\documentstyle[12pt,twoside,fleqn,epsfig,espcrc1]{article} 

\begin{document}

\title{Overview of $\bar K$-Nuclear Quasi-Bound States\footnote{Plenary talk 
given at Few-Body 18, Santos, Brazil, August 2006, to be published in 
Nucl. Phys. A}} 

\author{Avraham Gal\address{Racah Institute of Physics,
The Hebrew University, Jerusalem 91904, Israel}} 

\maketitle 

\begin{abstract}

Experimental evidence for $\bar K$-nuclear quasi-bound states is 
briefly reviewed. Theoretical and phenomenological arguments for 
and against the existence of such states are considered, based on 
constructing $\bar K$-nuclear optical potentials from various sources. 
Results of RMF calculations that provide a lower limit of 
$\Gamma_{\bar K} \sim 50 \pm 10$ MeV on the width of $\bar K$-nuclear 
quasi-bound states are discussed. 

\end{abstract}

\section{INTRODUCTION} 

The $\bar K$-nucleus interaction near threshold is strongly attractive and 
absorptive as suggested by fits to the strong-interaction shifts and widths 
of $K^-$-atom levels~\cite{BFG97}. 
Global fits yield `deep' optical potentials with 
Re~$V_{\bar K}(\rho_0) \sim -(150-200)$ MeV~\cite{FGB93,FGB94,FGM99,MFG06}, 
whereas more theoretically inclined studies that fit the low-energy 
$K^-p$ reaction data, including the $I=0$ quasi-bound state $\Lambda(1405)$ 
as input for constructing density dependent optical potentials, suggest 
relatively `shallow' potentials with Re~$V_{\bar K}(\rho_0) \sim -(40-60)$ 
MeV~\cite{SKE00,ROs00,CFG01}. The issue of depth of the attractive 
$\bar K$-nucleus potential is discussed in Section \ref{sec:pot}. 

\begin{figure}[th]
\centerline{\psfig{file=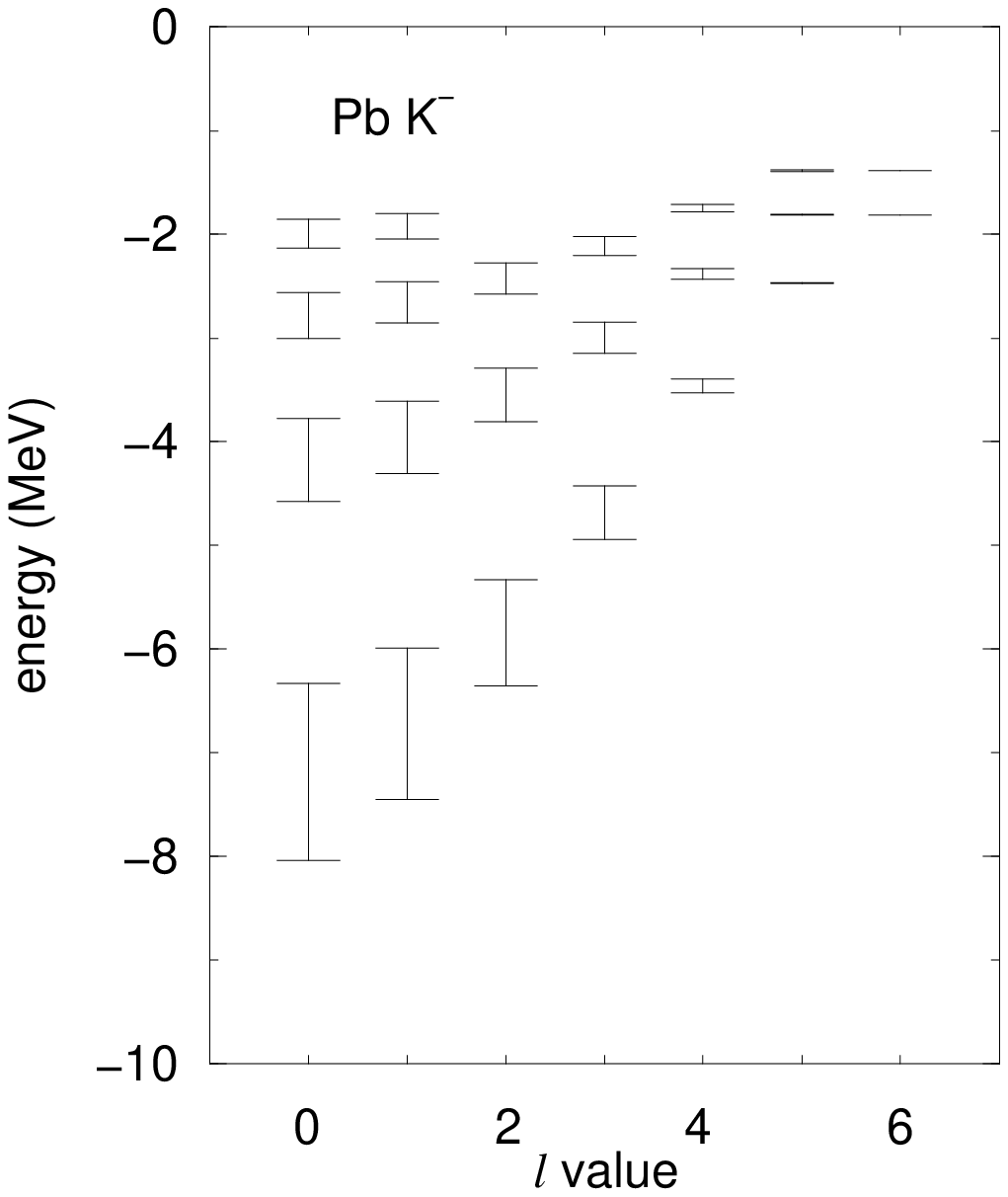,width=7.0cm}
\hspace*{3mm} 
\psfig{file=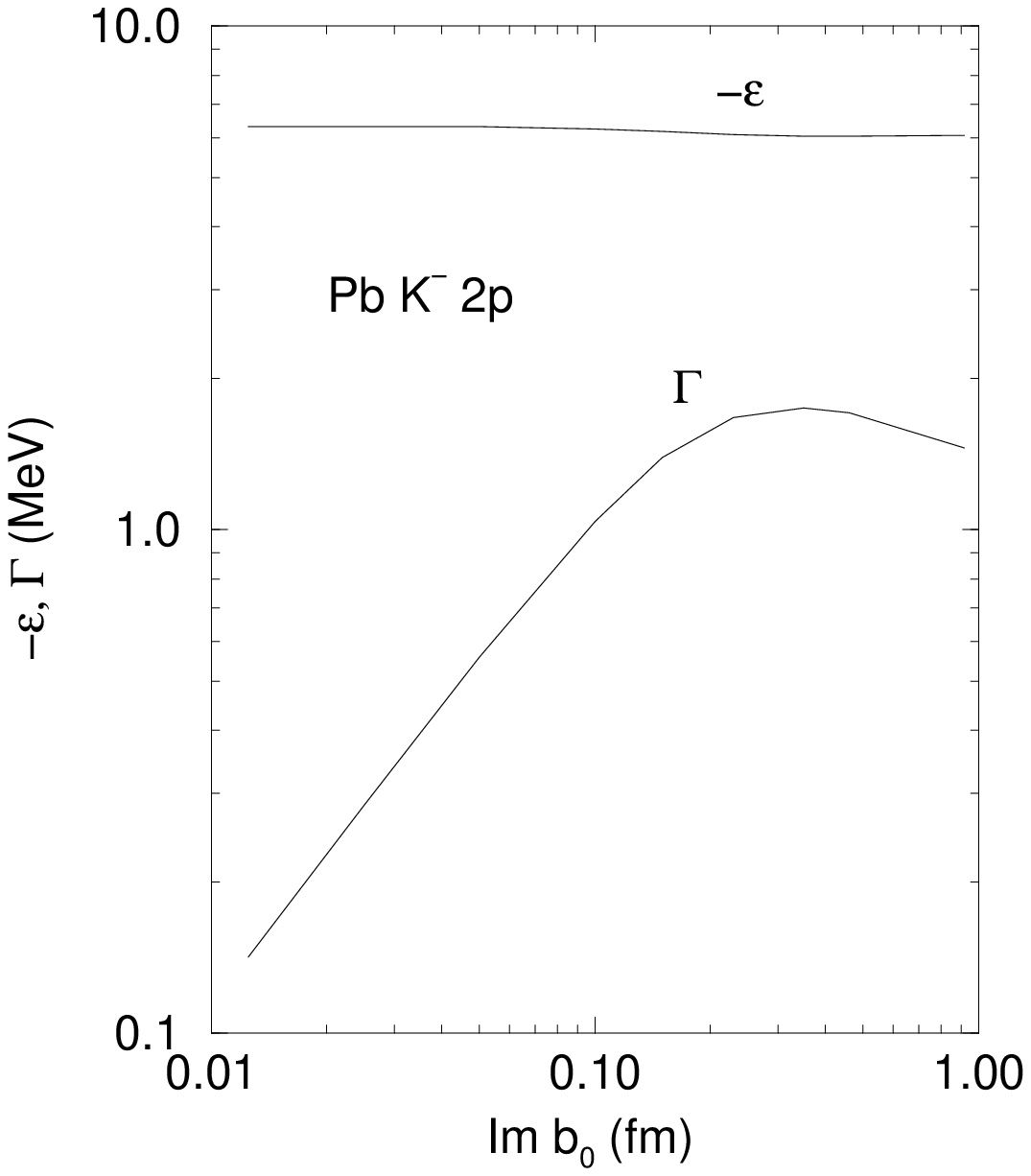,width=7.0cm}
}
\vspace*{8pt} 
\caption{Left: calculated $K^-$ `deeply bound' atomic states 
in $^{208}$Pb. Right: saturation of width $\Gamma$ for the $2p$ 
$K^-$-$^{208}$Pb atomic state as function of Im~$b_0$, 
for Re~$b_0 = 0.62$~fm. \label{fig:KPb}}
\end{figure}

Paradoxically, due to the strong (absorptive) Im~$V_{\bar K}$, 
relatively narrow $K^-$~deeply bound {\it atomic} states are expected  
to exist~\cite{FGa99a}, independently of the size of Re~$V_{\bar K}$.  
Figure~\ref{fig:KPb} from Ref.~\cite{FGa99b} shows on the left-hand side 
a calculated spectrum of $K^-$ atomic states in $^{208}$Pb where, 
in particular, all the circular states below the $7i~(l=6)$ state are 
not populated by X-ray transitions due to the strong $K^-$-nuclear absorption, 
and on the right-hand side it demonstrates saturation of the $2p$ atomic-state 
width as a function of the absorptivity parameter Im~$b_0$ of $V_{\bar K}$. 
The physics behind is that a strong Im~$V_{\bar K}$ 
acts repulsively, suppressing the {\it atomic} wavefunction in the region of 
overlap with Im~$V_{\bar K}$. The calculated width of the `deeply bound' 
atomic $1s$ and $2p$ is less than 2 MeV, also confirmed by the calculation 
of Ref.~\cite{BGN00}, calling for experimental ingenuity 
how to form these levels selectively by a non-radiative process~\cite{FGa99c}. 

This saturation mechanism does not hold for $\bar K$-nuclear states which 
retain very good overlap with the potential. Hence, the questions to ask 
are (i) whether it is possible at all to bind {\it strongly} $\bar K$ mesons 
in nuclei, and (ii) are such quasi-bound states sufficiently narrow to allow 
observation and identification? 
The first question was answered favorably by Nogami~\cite{Nog63} as early 
as 1963 arguing that the $K^-pp$ system could acquire about 10 MeV 
binding in its $I=1/2$ state. Yamazaki and Akaishi, within a single-channel 
$K^-pp$ calculation~\cite{YAk02}, reported a binding energy 
$B \sim 50$~MeV and width $\Gamma \sim 60$~MeV; however, 
the recent coupled-channel $\bar K NN - \pi \Sigma N$ Faddeev calculation for 
$K^-pp$ by Shevchenko et~al.~\cite{SGM06} finds a substantially broader 
$I=1/2$ state ($B \sim 60$~MeV, $\Gamma \sim 100$~MeV). 

The current experimental and 
theoretical interest in $\bar K$-nuclear bound states was triggered back in 
1999 by the suggestion of Kishimoto~\cite{Kis99} to look for such states 
in the nuclear reaction $(K^{-},p)$ in flight, and by Akaishi and 
Yamazaki~\cite{AYa99,AYa02} who suggested to look for a $\bar K NNN$ $I=0$ 
state bound by over 100 MeV for which the main $\bar K N \to \pi \Sigma$ 
decay channel would be kinematically closed.{\footnote{Wycech had conjectured 
that the width of such states could be as small as 20 MeV~\cite{Wyc86}.}} 
Some controversial evidence for relatively narrow states was presented 
initially in $(K^{-}_{\rm stop},n)$ and $(K^{-}_{\rm stop},p)$ reactions on 
$^4$He (KEK-PS E471)~\cite{ISB03,SBF04} but has just been 
withdrawn (KEK-PS E549/570)~\cite{Iwa06}. $\bar K$-nuclear states were also 
invoked to explain few weak irregularities in the neutron spectrum of the 
$(K^{-},n)$ in-flight reaction on $^{16}$O (BNL-AGS, parasite 
E930~\cite{KHA03}), but subsequent $(K^{-},n)$ and $(K^{-},p)$ reactions on 
$^{12}$C at $p_{\rm lab}=1$~GeV/c (KEK-PS E548~\cite{Kis06}) have not 
disclosed any peaks beyond the appreciable strength observed below the 
$\bar K$-nucleus threshold. Ongoing experiments by the FINUDA spectrometer 
collaboration at DA$\Phi$NE, Frascati, already claimed evidence for 
a relatively broad $K^- pp$ deeply bound state ($B \sim 115$~MeV) in 
$K^{-}_{\rm stop}$ reactions on Li and $^{12}$C, by observing back-to-back 
$\Lambda p$ pairs from the decay $K^-pp\to\Lambda p$~\cite{ABB05}, but these 
pairs could more naturally arise from conventional absorption processes at 
rest when final-state interaction is taken into account~\cite{MOR06}. 
Indeed, the $K^-_{\rm stop}pn\to \Sigma^- p$ reaction on $^6$Li has also 
been recently observed~\cite{ABB06}. Another recent 
claim for a very narrow and deep $K^- pp$ state ($B \sim 160$~MeV, 
$\Gamma \sim 30$~MeV) is also based on observing decay $\Lambda p$ pairs, 
using $\bar p$ annihilation data on $^4$He from the OBELIX spectrometer 
at LEAR, CERN~\cite{Bre06}. One cannot rule out that the $\Lambda p$ pairs 
assigned in the above analyses to $K^-pp$ decay in fact result from 
nonmesonic decays of different clusters, say the $\bar K NNN$ $I=0$ 
quasi-bound state. A definitive identification may only be reached through 
a fully exclusive formation analysis, such as the one scheduled for 
J-PARC~\cite{Nag06}: 
\begin{equation} 
\label{eq:nag}
K^-~+~^3{\rm He}~ \to ~ (K^- pp)~+~n \, . 
\end{equation} 
Finally, preliminary evidence 
for the $\bar K NNN$ $I=0$ state ($B \sim 50-60$~MeV, $\Gamma \sim 35$~MeV) 
has been recently presented by the FINUDA collaboration on $^6$Li by observing 
back-to-back $\Lambda d$ pairs~\cite{Pia06}. 
It is clear that the issue of $\bar K$ nuclear 
states is far yet from being experimentally resolved 
and more dedicated, systematic searches are necessary. 
 
It is interesting then to study theoretically $\bar K$ nuclear quasi-bound 
states in the range of binding energy $B_{\bar K} \sim 100 - 200$ MeV, 
in particular the width anticipated for such deeply bound states. 
This is described in Section~\ref{sec:RMF} within the relativistic mean field 
(RMF) model for a system of nucleons and one $\bar K$ meson interacting 
through the exchange of scalar ($\sigma$) and vector 
($\omega,\rho$) boson fields which are treated in the 
mean-field approximation~\cite{MFG06,MFG05}.

\section{$\bar K$-NUCLEUS POTENTIALS} 
\label{sec:pot} 
\subsection{Chirally motivated models} 
\label{sec:shallow} 

The Born approximation for $V_{\bar K}$ due to the 
leading-order Tomozawa-Weinberg (TW) vector term of the chiral effective 
Lagrangian \cite{WRW97} yields a sizable attraction: 
\begin{equation} 
\label{eq:chiral} 
V_{\bar K}=-\frac{3}{8f_{\pi}^2}~\rho\sim -55~\frac{\rho}{\rho_0}~~{\rm MeV} 
\end{equation} 
for $\rho _0 = 0.16$ fm$^{-3}$, where $f_{\pi} \sim 93$ MeV is the 
pseudoscalar meson decay constant. Iterating the TW term plus 
next-to-leading-order terms, 
within an {\it in-medium} coupled-channel approach constrained 
by the $\bar K N - \pi \Sigma - \pi \Lambda$ data near the 
$\bar K N$ threshold, roughly doubles this $\bar K$-nucleus attraction. 
It is found (e.g. Ref. \cite{WKW96}) that the $\Lambda(1405)$ quickly 
dissolves in the nuclear medium at low density, so that 
the repulsive free-space scattering length $a_{K^-p}$, as function of 
$\rho$, becomes {\it attractive} well below $\rho _0$. 
Since the attractive $I=1$ $a_{K^-n}$ is only weakly density 
dependent, the resulting in-medium $\bar K N$ isoscalar scattering length 
$b_0(\rho)={\frac{1}{2}}(a_{K^-p}(\rho)+a_{K^-n}(\rho)$) translates into 
a strongly attractive $V_{\bar K}$: 
\begin{equation} 
\label{eq:trho} 
V_{\bar K}(r) = -{\frac{2\pi}{\mu_{KN}}}~b_0(\rho)~\rho(r)~, 
~~~~{\rm Re}~V_{\bar K}(\rho_0) \sim -110~{\rm MeV}\,. 
\end{equation} 
However, when $V_{\bar K}$ is calculated {\it self consistently}, 
namely by including $V_{\bar K}$ in the propagator $G_0$ used in the 
Lippmann-Schwinger equation determining $b_0(\rho)$, one obtains 
Re~$V_{\bar K}(\rho_0)\sim -(40 - 60)$~MeV~\cite{SKE00,ROs00,CFG01}. 
The main reason for this weakening of $V_{\bar K}$, 
approximately back to that of Eq.~(\ref{eq:chiral}), 
is the strong absorptive effect which $V_{\bar K}$ exerts within $G_0$ to 
suppress the higher Born terms of the $\bar K N$ TW potential. 

\begin{figure}[th] 
\centerline{\psfig{file=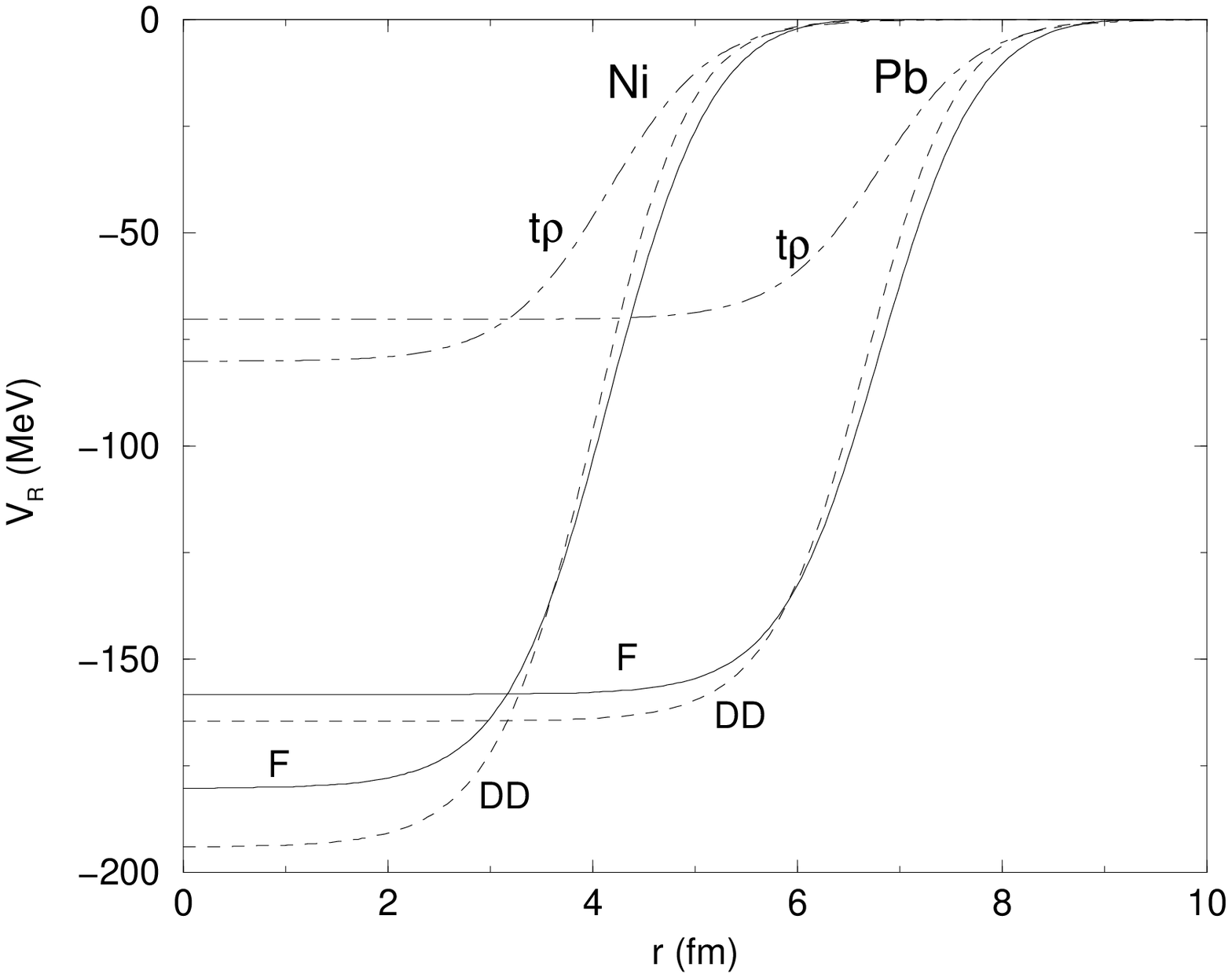,width=8.0cm}
\hspace*{3mm}
\psfig{file=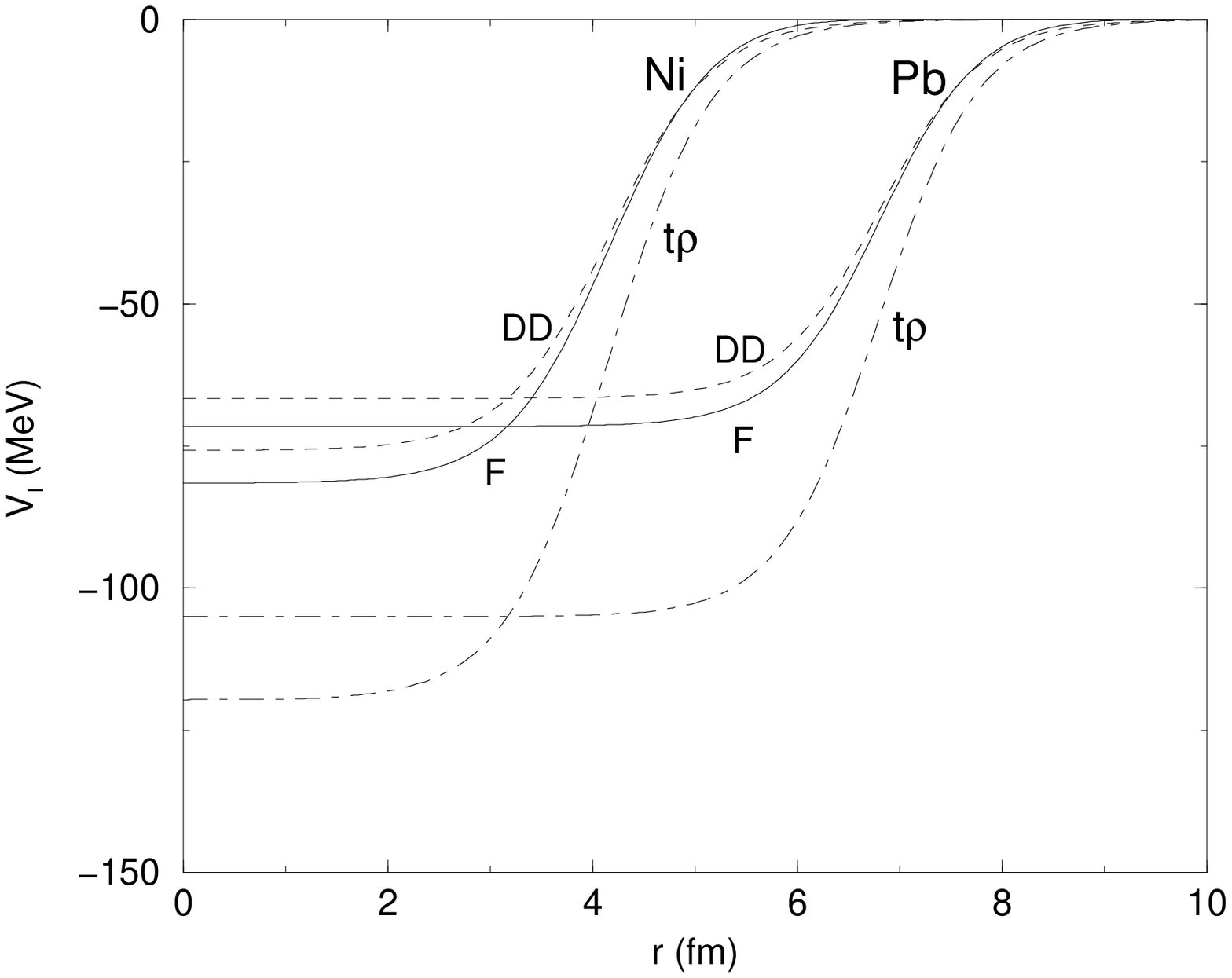,width=8.0cm}
}
\vspace*{8pt}
\caption{Real part (left) and imaginary part (right) of the 
$\bar K$-nucleus potential for $^{58}$Ni and $^{208}$Pb, 
obtained in a global fit to $K^-$-atom data, for a $t\rho$ 
potential, for the DD potential~{\protect \cite{BFG97}} and for potential 
F~{\protect \cite{MFG06}}, see text.} 
\label{fig:V} 
\end{figure}

\subsection{Fits to $K^-$-atom data} 
\label{sec:deep} 

The $K^-$-atom data used in global fits~\cite{BFG97} span a range of nuclei 
from $^7$Li to $^{238}$U, with 65 level-shift and -width data points. 
Figure \ref{fig:V} shows fitted $\bar K$-nucleus potentials for $^{58}$Ni 
and $^{208}$Pb, for a $t\rho$ potential with a complex strength $t$, and for 
two density-dependent potentials marked by DD and F, also fitted to the same 
data. For the real part, the depth of the $t\rho$ potential is about 
$70-80$~MeV, whereas the density-dependent potentials are considerably 
deeper, $150-200$~MeV. These latter potentials yield substantially lower 
$\chi ^2$ values of 103 and 84, respectively, than the value 129 for the 
$t\rho$ potential. We note that, 
although the two density-dependent potentials have very different 
parameterizations, the resulting potentials are quite similar. 
In particular, the shape of potential F departs
appreciably from  $\rho (r)$ for $\rho (r)/\rho_0 \leq 0.2$, where the
physics of the $\Lambda(1405)$ still plays some role. 
The density dependence of the potential F is qualitatively similar to that 
of Eq.~(\ref{eq:trho}), by far providing the best fit ever reported for 
a global $K^-$-atom data fit. 

\subsection{Further considerations}
\label{sec:add}

Additional considerations for estimating  $V_{\bar K}$ are listed below.  
 
(i) QCD sum-rule estimates~\cite{Dru06} for vector (v) and scalar (s) 
self-energies: 
\begin{eqnarray} 
\label{eq:QCDv} 
\Sigma_v(\bar K) &\sim & -\frac{1}{2}~\Sigma_v(N)~\sim~
-\frac{1}{2}~(200)~{\rm MeV}~ =~-100~{\rm MeV}\,,\\ 
\Sigma_s(\bar K) &\sim & \frac{m_s}{M_N}~\Sigma_s(N)~\sim~
\frac{1}{6}~(-300)~{\rm MeV}~ =~ -50~{\rm MeV}\, ,
\label{eq:QCDs}
\end{eqnarray}
where $m_s$ is the strange-quark (current) mass. The factor 1/2 in 
Eq.~(\ref{eq:QCDv}) is due to the one nonstrange antiquark in the $\bar K$ 
out of two possible, and the minus sign is due to $G$-parity going from 
quarks to antiquarks. This rough estimate gives then 
$V_{\bar K}(\rho_0) \sim -150$~MeV. The QCD sum-rule approach essentially 
refines the mean-field argument~\cite{SGM94,BRh96} 
\begin{equation} 
\label{eq:MF} 
V_{\bar K}(\rho_0)~\sim~\frac{1}{3}~(\Sigma_s(N)-\Sigma_v(N))~\sim~
-170~{\rm MeV}\,,
\end{equation} 
where the 1/3 factor is again due to the one nonstrange antiquark in the 
$\bar K$, but here with respect to the three nonstrange quarks of the nucleon. 

(ii) The ratio of $K^-/K^+$ production cross sections in nucleus-nucleus and
proton-nucleus collisions near threshold, measured by the Kaon spectrometer 
(KaoS) at SIS, GSI, gives clear evidence for attractive $V_{\bar K}$, 
estimated~\cite{SBD06} as $V_{\bar K}(\rho_0) \sim -80$~MeV by relying on 
BUU transport calculations normalized to the value $V_K(\rho_0) \sim +25$~MeV. 
Since the BUU calculations apparently disregard $\bar K NN \to YN$ absorption,
a deeper $V_{\bar K}$ may follow when nonmesonic absorption is included.

\section{RMF DYNAMICAL CALCULATIONS} 
\label{sec:RMF} 

\subsection{$\bar K$-nucleus RMF methodology} 

In this model, expanded in Ref. \cite{MFG06}, the (anti)kaon interaction with 
the nuclear medium is incorporated by adding to ${\cal L}_N$ the Lagrangian 
density ${\cal L}_K$: 
\begin{equation}
\label{eq:Lk}
{\cal L}_{K} = {\cal D}_{\mu}^*{\bar K}{\cal D}^{\mu}K -
m^2_K {\bar K}K
- g_{\sigma K}m_K\sigma {\bar K}K\; .
\end{equation} 
The covariant derivative
${\cal D_\mu}=\partial_\mu + ig_{\omega K}{\omega}_{\mu}$ describes
the coupling of the (anti)kaon to the vector meson $\omega$.
The coupling of the (anti)kaon to the isovector $\rho$ meson was 
neglected, a good approximation for the light $N=Z$ nuclei. 
The $\bar K$ meson induces additional source terms in the equations of motion 
for the meson fields $\sigma$ and $\omega_0$. It thus affects the scalar 
$S = g_{\sigma N}\sigma$ and the vector $V = g_{\omega N}\omega_0$ potentials 
which enter the Dirac equation for nucleons, and this leads to rearrangement 
or polarization of the nuclear core. Furthermore, in the Klein-Gordon 
equation satisfied by the $\bar K$, the scalar $S = g_{\sigma K}\sigma$ and 
the vector $V = -g_{\omega K}\omega_0$ potentials become 
{\it state dependent} through the {\it dynamical} density dependence of the 
mean-field potentials $S$ and $V$, as expected from a RMF calculation. 
The $\bar K$ potential was made complex by adding ${\rm Im}~V_{\rm opt}$ 
in a $t\rho$ form fitted to the $K^-$ atomic data~\cite{FGM99}. 
From phase-space considerations, ${\rm Im}~V_{\rm opt}$ was then multiplied 
by a suppression factor, taking into account the binding energy 
of the antikaon for the initial decaying state, and assuming two-body 
final-state kinematics for the decay products in the $\bar K N \to \pi Y$ 
mesonic modes and in the $\bar K NN \to Y N$ nonmesonic modes.  

The coupled system of equations was solved self-consistently. 
To give rough idea, whereas the static calculation gave $B_{K^-} = 132$~MeV
for the $K^-$ $1s$ state in $^{12}$C, using the values 
$g^{\rm atom}_{\omega K},~g^{\rm atom}_{\sigma K}$ corresponding to the 
$K^-$-atom fit, the dynamical calculation gave $B_{K^-} = 172$~MeV for this 
same state. In order to produce different values of binding energies, 
$g_{\sigma K}$ and $g_{\omega K}$ were varied in given intervals of physical 
interest. 

\subsection{Binding energies and widths}

Beginning approximately with $^{12}$C, the following conclusions may be drawn: 

(i) For given values of $g_{\sigma K},g_{\omega K}$, the $\bar K$ binding 
energy $B_{\bar K}$ saturates, except for a small increase due to the 
Coulomb energy (for $K^-$). 

(ii) The difference between the binding energies calculated dynamically and 
statically, $B_{\bar K}^{\rm dyn} - B_{\bar K}^{\rm stat}$, is substantial 
in light nuclei, increasing with $B_{\bar K}$ for a given value of $A$, and 
decreasing monotonically with $A$ for a given value of $B_{\bar K}$. 
It may be neglected only for very heavy nuclei. The same holds for the 
nuclear rearrangement energy $B_{\bar K}^{\rm s.p.} - B_{\bar K}$ which is 
a fraction of $B_{\bar K}^{\rm dyn} - B_{\bar K}^{\rm stat}$. 
 
\begin{figure}[th]
\centerline{\psfig{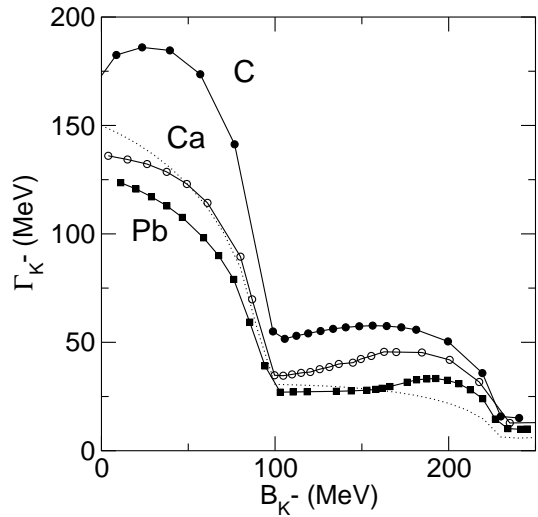}}
\vspace*{8pt} 
\caption{Dynamically calculated widths of the $1s$ $K^-$-nuclear state
in $^{~~12}_{K^-}$C, $^{~~40}_{K^-}$Ca and $^{~~208}_{K^-}$Pb 
as function of the $K^-$ binding energy for nonlinear RMF models. 
The dotted line is for a static nuclear-matter calculation with 
$\rho_0=0.16~{\rm fm}^{-3}$, see text.}
\label{fig:Gamma}  
\end{figure}

(iii) The width $\Gamma_{\bar K}(B_{\bar K})$ decreases monotonically 
with $A$, according to Fig.~\ref{fig:Gamma} 
which shows calculated 
widths $\Gamma_{K^-}$ as function of the binding energy $B_{K^-}$ for 
$1s$ states in $^{~~12}_{K^-}$C and $^{~~40}_{K^-}$Ca, using the nonlinear 
NL-SH version~\cite{SNR93} of the RMF model, and in 
$^{~~208}_{K^-}$Pb using the NL-TM1 version~\cite{STo94}. 
The dotted line shows the static `nuclear-matter' limit 
corresponding to the $K^-$-atom fitted value ${\rm Im}~b_0=0.62$ fm and for 
$\rho(r)=\rho_0=0.16$ fm$^{-3}$, using the same phase-space suppression factor 
as in the `dynamical' calculations. It is clearly seen that the functional 
dependence $\Gamma_{K^-}(B_{K^-})$ follows the shape of the dotted line. 
This dependence is due primarily to the binding-energy
dependence of the suppression factor $f$ which falls off rapidly until
$B_{K^-} \sim 100$~MeV, where the dominant
$\bar K N \rightarrow \pi \Sigma$ gets switched off, and then stays
rather flat in the range $B_{K^-} \sim 100 - 200$~MeV where the width is 
dominated by the $\bar K NN \to YN$ absorption modes. The widths
calculated in this range are considerably larger than given by the dotted 
line (except for Pb in the range $B_{K^-} \sim 100 - 150$~MeV) due 
to the dynamical nature of the RMF calculation, whereby the nuclear density 
is increased by the polarization effect of the $K^-$. 
Adding perturbatively the residual width neglected in this calculation, 
partly due to the $\bar K N \to \pi \Lambda$ secondary mesonic decay channel, 
a representative value for a lower limit $\Gamma_{\bar K} \sim 50 \pm 10$~MeV 
holds in the binding energy range $B_{K^-} \sim 100 - 200$~MeV.

\subsection{Nuclear polarization effects} 

\begin{figure}[th]
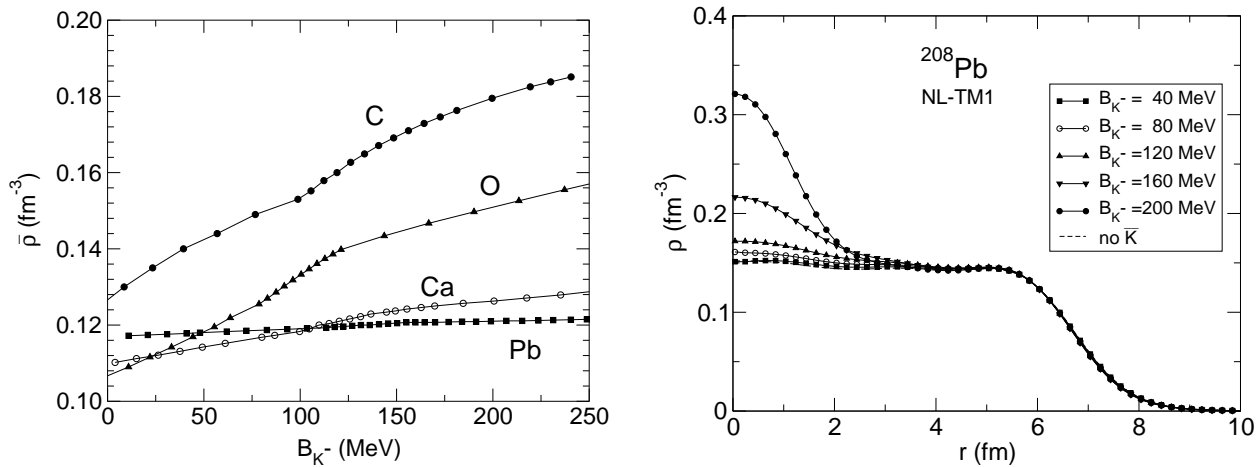

\centerline{\psfig{file=K05fig6.eps,width=8.0cm} 
\hspace*{3mm}
\psfig{file=K06PbTM1rho.eps,width=8.0cm}
}
\vspace*{8pt}
\caption{Calculated average nuclear density $\bar \rho$ for 
$^{~~12}_{K^-}$C, $^{~~16}_{K^-}$O, $^{~~40}_{K^-}$Ca and $^{~~208}_{K^-}$Pb 
(left) as function of $B_{K^-}(1s)$ for the NL-SH RMF 
model~{\protect \cite{SNR93}}, and nuclear density $\rho$ 
of $^{~208}_{K^-}$Pb (right) for several $B_{K^-}(1s)$ values, using the 
NL-TM1 RMF model~{\protect \cite{STo94}}.} 
\label{fig:rho}
\end{figure} 

Figure \ref{fig:rho} shows on the left-hand side the calculated average 
nuclear density $\bar \rho = \frac{1}{A}\int\rho^2d{\bf r}$ as a function 
of $B_{K^-}$ for $K^-$ nuclear $1s$ states across the periodic table. 
In the light $K^-$ nuclei, $\bar \rho$ increases substantially with $B_{K^-}$ 
to values about 50\% higher than without the $\bar K$. 
The increase of the central nuclear densities is even bigger, 
as demonstrated on the right-hand side for a family of $1s$ $K^-$ states 
in $^{~208}_{K^-}$Pb, but is confined to a small region subsiding almost 
completely by $r=2$~fm, well within the nuclear volume. 
As a result, ${\bar \rho}(^{~208}_{K^-}{\rm Pb})$ on the left-hand side of 
Fig.~\ref{fig:rho} is only weakly enhanced as function of $B_{K^-}$.

\section{CONCLUSIONS} 

In this talk I reviewed the phenomenological and theoretical evidence for 
a substantially attractive $\bar K$-nucleus interaction potential, from 
a `shallow' potential of depth $40-60$ MeV to a `deep' potential of depth 
$150-200$ MeV at nuclear-matter density. I then reported on recent 
{\it dynamical} calculations~\cite{MFG06} for deeply quasi-bound 
$K^-$~nuclear states across the periodic table. Substantial polarization 
of the core nucleus was found in light nuclei, but the `high' densities 
reached are considerably lower than those found in the few-body calculations 
due to Akaishi, Yamazaki and collaborators~\cite{YAk02,AYa02,DHA04a,DHA04b}. 
An almost universal dependence of $\bar K$ widths on the binding energy was 
found, for a given nucleus, reflecting the phase-space suppression factor on 
top of the increase provided by the density of the compressed nuclear core. 
The present results place a lower limit $\Gamma_{\bar K} \sim 50 \pm 10$~MeV 
which is particularly useful for $\bar K$ binding energies exceeding 100~MeV.

\section*{ACKNOWLEDGMENTS}

I wish to thank my collaborators Eli Friedman, Ji\v{r}\'{\i} Mare\v{s} 
and Nina Shevchenko, as well as Tullio Bressani, Evegny Drukarev, 
Tomofumi Nagae and Wolfram Weise for stimulating discussions, and Lauro Tomio 
and the co-organizers of FB18 for their kind hospitality and support. Special 
thanks go to Wolfram Weise for his kind hospitality at the TU Munich where 
this report was prepared under the support of the Alexander von Humboldt 
Foundation. This work is supported in part by the Israel Science Foundation, 
Jerusalem, grant 757/05.


\begin{thebibliography}{bit99}

\bibitem{BFG97} C.J. Batty, E. Friedman, A. Gal, Phys. Rep. 287 (1997) 385.

\bibitem{FGB93} E. Friedman, A. Gal, C.J. Batty, Phys. Lett. B 308 (1993) 6. 

\bibitem{FGB94} E. Friedman, A. Gal, C.J. Batty, Nucl. Phys. A 579 (1994) 518.

\bibitem{FGM99} E. Friedman, A. Gal, J. Mare\v{s}, A. Ciepl\'y, Phys. Rev. C 
60 (1999) 024314.

\bibitem{MFG06} J. Mare\v{s}, E. Friedman, A. Gal, Nucl. Phys. A 770 (2006) 84. 

\bibitem{SKE00} J. Schaffner-Bielich, V. Koch, M. Effenberger, Nucl. Phys. A 
669 (2000) 153. 

\bibitem{ROs00} A. Ramos, E. Oset, Nucl. Phys. A 671 (2000) 481. 

\bibitem{CFG01} A. Ciepl\'y, E. Friedman, A. Gal, J. Mare\v{s}, Nucl. Phys. A 
696 (2001) 173.

\bibitem{FGa99a} E. Friedman, A. Gal, Phys. Lett. B 459 (1999) 43. 

\bibitem{FGa99b} E. Friedman, A. Gal, Nucl. Phys. A 658 (1999) 345. 

\bibitem{BGN00} A. Baca, C. Garc\'{\i}a-Recio, J. Nieves, Nucl. Phys. A 673 
(2000) 335. 

\bibitem{FGa99c} E. Friedman, A. Gal, in: S. Bianconi, et al. (Eds.), 
Proc. III Int. DA$\Phi$NE Workshop, Frascati Physics Series vol. XVI, LNF, 
Frascati, 1999, pp. 677-684. 

\bibitem{Nog63} Y. Nogami, Phys. Lett. 7 (1963) 288.

\bibitem{YAk02} T. Yamazaki, Y. Akaishi, Phys. Lett. B 535 (2002) 70. 

\bibitem{SGM06} N.V. Shevchenko, A. Gal, J. Mare\v{s}, submitted for 
publication, arXiv:nucl-th/0610022.  

\bibitem{Kis99} T. Kishimoto, Phys. Rev. Lett. 83 (1999) 4701.

\bibitem{AYa99} Y. Akaishi, T. Yamazaki, in: S. Bianconi, et al. (Eds.), 
Proc. III Int. DA$\Phi$NE Workshop, Frascati Physics Series vol. XVI, LNF, 
Frascati, 1999, pp. 59-74. 

\bibitem{AYa02} Y. Akaishi, T. Yamazaki, Phys. Rev. C 65 (2002) 044005. 

\bibitem{Wyc86} S. Wycech, Nucl. Phys. A 450 (1986) 399c. 

\bibitem{ISB03} M. Iwasaki, et al., nucl-ex/0310018; 
T. Suzuki, et al., Nucl. Phys. A 754 (2005) 375c.

\bibitem{SBF04} T. Suzuki, et al., Phys. Lett. B 597 (2004) 263. 

\bibitem{Iwa06} M. Iwasaki, plenary talk at HYP06, Mainz, October 2006. 

\bibitem{KHA03} T. Kishimoto, et al., Nucl. Phys. A 754 (2005) 383c.

\bibitem{Kis06} T. Kishimoto, private communication, September 2006. 

\bibitem{ABB05} M. Agnello, et al., Phys. Rev. Lett. 94 (2005) 212303. 

\bibitem{MOR06} V.K. Magas, E. Oset, A. Ramos, H. Toki, Phys. Rev. C 74 
(2006) 025206. 

\bibitem{ABB06} M. Agnello, et al., Nucl. Phys. A 775 (2006) 35. 

\bibitem{Bre06} T. Bressani, presented at HYP06, Mainz, October 2006.

\bibitem{Nag06} T. Nagae, plenary talk at HYP06, Mainz, October 2006. 

\bibitem{Pia06} S. Piano, presented at HYP06, Mainz, October 2006. 

\bibitem{MFG05} J. Mare\v{s}, E. Friedman, A. Gal, Phys. Lett. B 606 (2005) 
295. 

\bibitem{WRW97} T. Waas, M. Rho, W. Weise, Nucl. Phys. A 617 (1997) 449, 
and references therein. 

\bibitem{WKW96} T. Waas, N. Kaiser, W. Weise, Phys. Lett. B 379 (1996) 34. 

\bibitem{Dru06} E.G. Drukarev, private communication, May 2006.  

\bibitem{SGM94} J. Schaffner, A. Gal, I.N. Mishustin, H. St\"{o}cker, et al., 
Phys. Lett. B 334 (1994) 268. 

\bibitem{BRh96} G.E. Brown, M. Rho, Nucl. Phys. A 596 (1996) 503. 

\bibitem{SBD06} W. Scheinast, et al., Phys. Rev. Lett. 96 (2006) 072301, 
and references therein.

\bibitem{SNR93} M.M. Sharma, M.A. Nagarajan, P. Ring, Phys. Lett. B 312 
(1993) 377.

\bibitem{STo94} Y. Sugahara, H. Toki, Nucl. Phys. A 579 (1994) 557. 

\bibitem{DHA04a} A. Dot\'e, H. Horiuchi, Y. Akaishi, T. Yamazaki, 
Phys. Lett. B 590 (2004) 51. 

\bibitem{DHA04b} A. Dot\'e, H. Horiuchi, Y. Akaishi, T. Yamazaki, 
Phys. Rev. C 70 (2004) 044313. 

\end{thebibliography}
\end{document}